\documentclass[preprint,pra,aps,showpacs,amsmath]{revtex4}

\usepackage{epsfig}

\begin{document}

\author{Nikolai G.\ Kalugin}
\affiliation{
Institute for Quantum Studies and Department of Physics, 
%$^2$Chem., and Elec. Eng., \\
Texas A\&M University,
%College Station, 
TX 77843
}
\author{Yuri V.\ Rostovtsev}
\affiliation{
Institute for Quantum Studies and Department of Physics, 
%$^2$Chem., and Elec. Eng., \\
Texas A\&M University,
%College Station, 
TX 77843
}

\author{Marlan\ O.\ Scully}
\affiliation{
Institute for Quantum Studies and Depts. of Physics, 
$^2$Chem., and Elec. Eng., \\
Texas A\&M University,
%College Station, 
TX 77843
}
\affiliation{
Princeton Inst. for Material Sci. and Department of Mechanical \& Aerospace 
Engineering, Princeton University, NJ 08544}

\title{Generation of strong short coherent terahertz pulses in 
ladder-Lambda and double-Lambda systems}

\begin{abstract}
We show that coherently driven atomic or molecular media  
potentially yield strong controllable 
short pulses of THz radiation. The method
is based on excitation of maximal quantum coherence in a gas medium  
by optical pulses and coherent scattering of infra-red radiation 
to produce pulses of THz radiation. The pulses have the 
energies range from several nJ 
to $\mu$J and time durations from several fs to ns 
at room temperature.
\end{abstract}
%\pacs{42.50.Gy}
\date{\today}
\maketitle

\newcommand{\ds}{\displaystyle}
\newcommand{\dd}{\partial}
\newcommand{\be}{\begin{equation}}
\newcommand{\ee}{\end{equation}}
\newcommand{\beq}{\begin{eqnarray}}
\newcommand{\eeq}{\end{eqnarray}}
\newcommand{\dt}{\ds\frac{\dd}{\dd t}}
\newcommand{\dz}{\ds\frac{\dd}{\dd z}}
\newcommand{\D}{\ds\left(\frac{\dd}{\dd t} + c \frac{\dd}{\dd z}\right)}

\newcommand{\w}{\omega}
\newcommand{\W}{\Omega}
\newcommand{\g}{\gamma}
\newcommand{\G}{\Gamma}
\newcommand{\E}{\hat E}
\newcommand{\s}{\sigma}
%\newcommand{\bra}{\langle}
%\newcommand{\ket}{\rangle}

%\section{Motivation of generation ultra-short THz pulses}

TeraHertz (THz) radiation 
%(a frequency interval from 0.1 to 10 THz,
%wavelengths from 3 mm to 30$\mu$m, wavenumbers  from 3 to 340 cm$^{-1}$, 
%photon energies from  0.4 to 40 meV), 
has unique potential for applications to material
diagnostics including semiconductors, chemical compounds, biomolecules and
biotissues \cite{1,2,5}. 
%
%The THz imaging and microscopy open new and 
%very attractive possibilities for medical diagnostics~\cite{2}.
%Ultrafast THz sources in combination with a high pulse magnetic 
%fields \cite{5},
%can trigger a real revolution in electron magnetic resonance
%spectroscopy techniques in many fields. 
%Potential applications of high
%intensity sources include driving new nonlinear phenomena in THz
%region.  
%
However, despite significant progress in recent
years~\cite{14,10,detectors}, 
the methods of generation of THz radiaiton are still less developed 
than in the visible and  
near-infrared regions. The search for efficient, high-power,
nonexpensive, portable, and room-temperature suitable 
methods of generation of THz
radiation is one of the main topics in modern optoelectronics and photonics. 
%Our approach to the problem is to take advantage of the dramatic enhancement 
%of the nonlinear response via quantum coherence
%\cite{eit,matsko,sokolov,harris,nazarkin,cars}. 

\begin{figure} %1
\center{
\includegraphics[width=16cm]{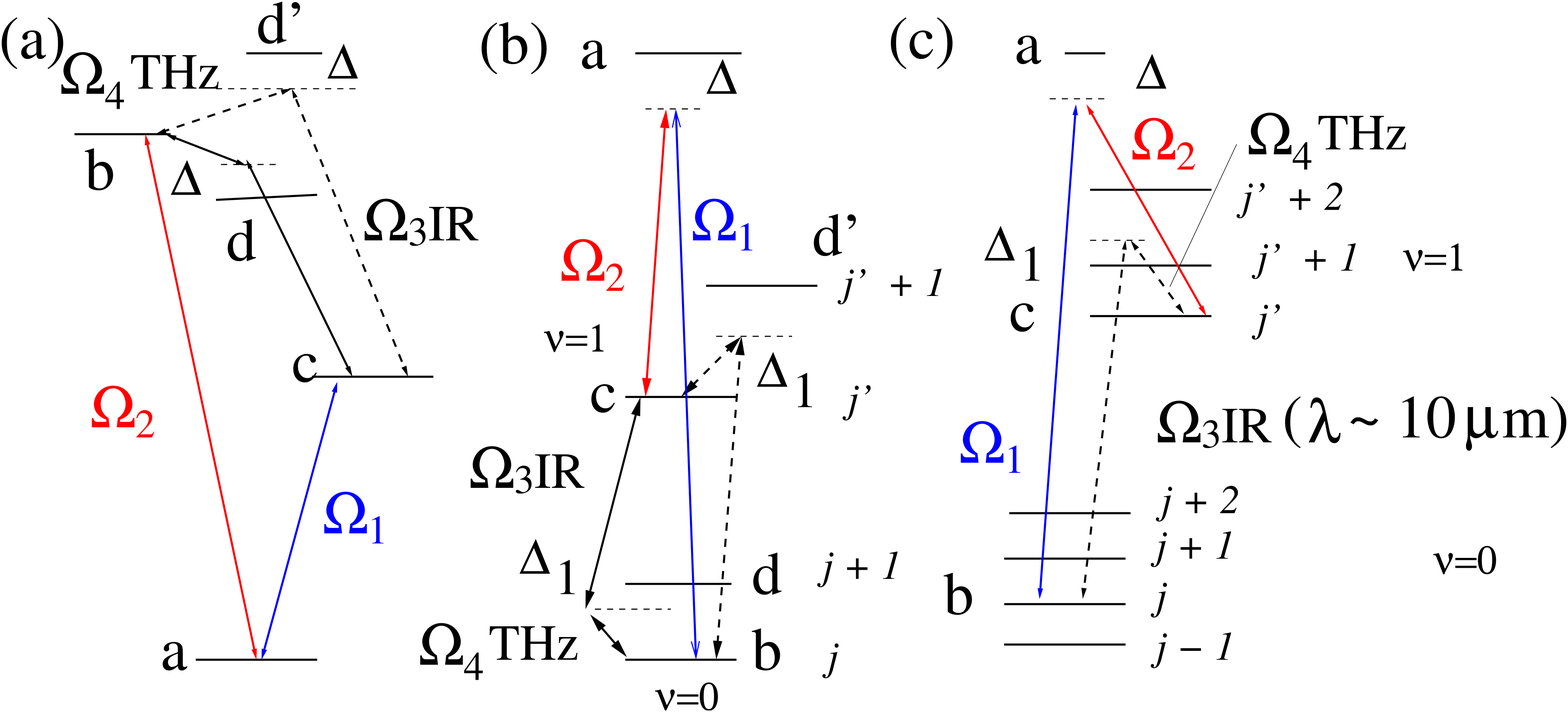}
}
\caption{\label{main-scheme1}
%
%Level schemes  for THz generation.
%$\W_1$ and $\W_2$ are optical fields in two-photon resonance to excite atomic
%or molecular coherence between levels $b$ and $c$. 
%$\W_3$ is the IR field, and $\W_4$ is generated THz radiation 
%(Double-$\Lambda$ shown by dashed lines, Ladder-$\Lambda$ by solid lines). 
%Atomic medium (a). For example, Rb levels are 
%$a=5S_{1/2}$, $b=10P_{1/2.3/2}$, $c=6P_{1/2.3/2}$, $d(d')=8D_{3/2,5/2}
%(9D_{3/2,5/2})$.  
%Molecular medium (b). Level $b$ is the ground state, vibrational quantum
%number $v=0$ and angular momentum $j$; $c$ is the level with $v=1$ and $j'$, 
%$d$ is the level, $v=0$ and $j+1$; $d'$ is the level $v=1$ and $j'+1$. 
%Levels of CH$_3$F shown in (c).
%
}
\end{figure}

In this Letter, we suggest a way to generate 
short pulses of coherent THz radiation with high efficiency
by using a coherently driven medium 
with excited maximal coherence induced by optical fields
\cite{eit,matsko,sokolov,harris,nazarkin,cars}. 
This idea can be implemented in several schemes shown 
in Fig.~\ref{main-scheme1} using atomic and molecular media 
involving rotational and vibrational molecular levels. 
Two optical fields with Rabi frequencies $\W_1$ and $\W_2$ 
propagate through a cell filled with a gas 
(the gases widely used for
generation of THz generation are methanol (CH$_3$OH), CH$_3$F, 
H$_2$F$_2$, CH$_3$Cl, 
etc. \cite{22,23,24}; simplified level structures are 
shown in Fig.~\ref{main-scheme1}) and prepare maximal 
coherence $\rho_{bc}$ between vibrational levels $b$ and $c$. 
Then, coherent scattering of IR radiation $\W_3$ produces 
THz radiation $\W_4$. 

The interaction Hamiltonian for the system is given by
\beq
V_I = -\hbar[\W_{2}e^{-i\w_{ac}t}|a\rangle\langle c|
+\W_{1}e^{-i\w_{ab}t}|a\rangle\langle b| + h.c.]\nonumber \\
-\hbar[\W_{3}e^{-i\w_{dc}t}|d\rangle\langle c|
+\W_{4}e^{-i\w_{db}t}|d\rangle\langle b| + h.c.]
\eeq
$\Omega_{i}=\wp_{i}{\cal E}_{i}/\hbar$ is the
Rabi frequency of the respective fields; 
$\wp_{ab}$ and $\wp_{ac}$ are the electrical dipole matrix elements 
between states $a$ and $b$, and $a$ and $c$; 
$\w_{ab}$ and $\w_{ac}$ are the frequencies of the electronic transitions;
$\w_{cd}$ and $\w_{db}$ are the frequencies of the vibrational 
and rotational transitions; ${\cal E}_{i}$ is the amplitude of the
respective laser field. 

The time-dependent density matrix equations are
\be
\frac{\partial{\rho}}{\partial{\tau}}
=-\frac{i}{\hbar}[H, \rho]-\frac{1}{2}(\Gamma\rho+\rho\Gamma),
\ee
where $\G$ is the relaxation matrix. 
%Solving these equations gives the time evolution of 
%the density matrix elements. 
A self-consistent system also includes 
the field propagation equations
%%
%\be
%\frac{\partial\W_1}{\partial{z}}=-\kappa_1\W_1-i\eta_1\rho_{ab},
%\;\;\;
%\frac{\partial\W_2}{\partial{z}}=-\kappa_2\W_2-i\eta_2\rho_{ac},
%\ee
%\be
%\frac{\partial\W_3}{\partial{z}}=-\kappa_3\W_3-i\eta_3\rho_{db},
%\;\;\;
%\frac{\partial\W_4}{\partial{z}}=-\kappa_4\W_4-i\xi\eta_4\rho_{dc},
%\ee
%%
\be
\frac{\partial\W_\alpha}{\partial{z}}=-\kappa_\alpha\W_\alpha-
i\xi\eta_\alpha\rho_\alpha,
\ee
where, index $\alpha=1,2,3,4$ indicates all fields and corresponding 
polarizations, 
$\eta_\alpha=\nu_\alpha N \wp_\alpha/(2\epsilon_0 c)$,  
%$\eta_2=\nu_2 N \wp_{ac}/(2\epsilon_0 c)$,
%$\eta_3=\nu_3 N \wp_{v}/(2\epsilon_0 c)$, and 
%$\eta_4=\nu_4 N \wp_{j}/(2\epsilon_0 c)$
is the corresponding coupling constant,
$\xi = \int  F_{nm}(x,y) {dxdy/S}$ 
is the filling factor of the TE$_{nm}$ mode 
of the waveguide, $F_{nm}(x,y)$ is the transverse field dependence~\cite{TEM}, 
$S$ is the area of the waveguide,   
$\nu_{1,2,3,4}$ are the frequencies of the optical, IR, and THz fields; 
$N$ is the density of medium,
$\epsilon_0$ is the permitivity of the vacuum, $c$ is the
speed of light in vacuum, and $\kappa_i$ are losses of the field 
in the cell because of scattering, diffraction, or 
nonresonant absorption. 
%For the THz field in free space the diffraction losses 
%given by $\kappa_4 = \lambda/D^2$ should be taken into account. 
To avoid diffraction losses the cell may be placed in the waveguide for THz.
%then the distribution of the field mode should be taken into account. 

%The above system can be solved in the ultra-short pulsed regime. 
As shown in \cite{harris,sokolov}, vibrational coherence $\rho_{cb}$
can be excited as much as $0.1-0.5$. The field $\W_4$ at the output of the
cell is given by 
%\be
%\W_4 = \xi\int_0^L dz e^{i(k_4 - \Delta k - k_3)z}{\eta\rho_{cb}\W_3^*
%\over\G_{db}}
%\ee
\be
\W_4 = \xi\int_0^L dz e^{i\delta kz}\eta\tau\rho_{cb}\W_3^* = \xi\ds
{\sin(\delta kL)\over
\ds(\delta kL)} 
\eta\tau\rho_{cb} L\W^*_3
\ee
where $L$ is the length of the cell, 
%Using a waveguide for the field $\W_4$ decreases diffraction losses 
%and leads to more efficient THz generation, 
%The cylindrical shape waveguide filled by gas (similar solution for 
%generation of ultrafast visible-UV pulses was used by authors of 
%[sokolov,f.e ]), or more sophisticated electrodynamical scheme 
%embedded into a gas cell. In the case of a simple cylindrical waveguide, 
%the expression for scattered filed will transform to  
%
%
%
%After integration 
%we obtain for scattered field $\W_4$ 
%\be
%\W_4 = \xi\ds
%{\sin(\delta kL)\over
%\ds(\delta kL)} 
%{\eta\rho_{cb} L\W^*_3\over\G_{db}}
%\ee
%where 
and the pre-factor describes phase-matching 
($\delta k = k_4 - k_3 - k_1 + k_2$), $\xi\simeq 1$.
% is the factor describing overlapping between mode of 
%the IR field and the THz radiation while propagating in the waveguide
%to decrease the diffraction losses for THz. 
 
\begin{figure*} %1
\center{
\includegraphics[width=18cm]{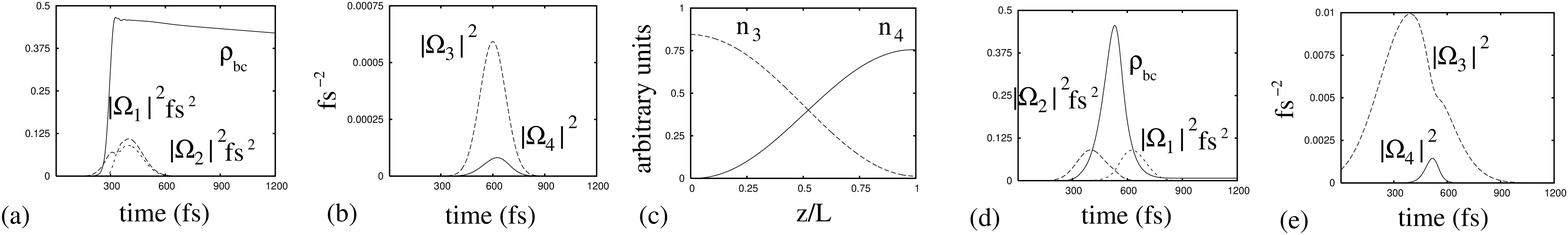}
}
\caption{\label{simulations}
%
%Results of simulations for double-$\Lambda$ scheme in CH$_3$F for two temporal
%scenarios. 
%Scenario 1. 
%Fields $\W_1$ and $\W_2$ and prepared by them coherence $\rho_{bc}$ 
%are shown in a).  Probe IR field $\W_3$ and generated THz field $\W_4$ are 
%shown in b).  Efficiency of nonlinear transformation is shown in c).  
%Scenario 2.
%Fields $\W_1$ and $\W_2$ in STIRAP configuration 
%and prepared by them coherence $\rho_{bc}$ 
%are shown in d).  Probe IR field $\W_3$ and generated THz field $\W_4$ are 
%shown in e).
%Parameters used in simulations are in the text.
%
}
\end{figure*}

The THz radiation can be generated in a cell with a waveguide 
filled with a atomic or molecular gas by applying 3 parallel coaxial beams 
(two visible pump beams to prepare vibrational coherence, 
and an IR probe beam to produce THz radiaiton).
There are two pulsed regime scenarios to generate THz radiation. 
First, we apply two short optical laser pulses 
(for example, femtosecond) to excite quantum coherence
~\cite{cars} and then a short IR pulse  
to obtain $\W_4$ as a result of coherent scattering of  
field $\W_3$ on this coherence (simulations are shown in 
Fig.~\ref{simulations}a,b,c). 
%\be
%\W_4 = \int^t_0 \rho_{cb}\exp[-\tilde\G (t-t')]\W_3(t')dt'
%\ee
%where $\tilde\G = \G_{cb} + |\W|^2/\G_{dc}$.

The second scenario allows us to obtain a pulse of THz radiation shorter than
the IR pulse by applying two optical pulse in 
the counter-intuitive pulse sequence \cite{STIRAP}. 
Large coherence is excited during stimulated adiabatic passage
(STIRAP), and the THz radiation occurs via coherent scattering during 
the time of the existence of maximal quantum coherence 
(simulations are shown in 
Fig.~\ref{simulations}d,e).
The efficiencies for both these processes are the same and given by 
\be
\epsilon = {I_4^L\over I_3^0} = 
\xi^2\mbox{sinc}^2(\delta kL)
\left({4\pi^2\wp_v\wp_j N\rho_{bc}\tau L
\over\lambda\hbar}
\right)^2,
\label{eff2}
\ee
where $I_3^0$ and $I_4^L$ are the intensities of IR at the input 
and THz radiation at the output of the cell
correspondingly, $\wp_v$ and $\wp_j$ are the dipole moments of transitions 
between vibrational and rotational levels correspondingly, 
$\tau$ is the duration of the THz pulse, $\lambda$ is the wavelength 
of THz radiaiton, and $L$ is the length of the cell.

In the case of methanol, 
the energy of the THz pulse can be as much as 6 nJ (using comercially 
available IR CO$_2$ laser, 6 kW, 15 ns. On the other hand, 
with a better source of IR radiation, pulses of THz radiation with 
energies of 10 $\mu$J can be obtained, 
for example, by using the CO$_2$ laser with 5 mJ energy and 
a duration of 500 ps \cite{co2}).  
The estimated efficiency can be as much as 
$\epsilon \simeq 1$ (for parameters $\tau = 10^{-12}$ s, $L=10$ cm, 
$\wp_v=\wp_j = 1D = 10^{-18}$ esu cm, $N=5\; 10^{16}$ cm$^{-3}$, 
$\lambda=100$ $\mu$m). 
Note here that Eq.(5) does not take into accounnt depletion of the probe
field. The overall efficiency have to
be estimated by Manley-Rowe relations which are different for different
proposed schemes. For example, for the double-$\Lambda$ scheme, it is   
$n_1 - n_2 + n_3 - n_4 = 0$, where $n_i$ is the number of photons in the $i$th
beam. The maximum $\epsilon$ one can obtain is given by $n_3 = n_4$. But 
for the ladder-$\Lambda$ scheme, $n_1 - n_2 - n_3 - n_4 = 0$ 
and $n_3$ and $n_4$ both grow during nonlinear transformation, efficiency as
defined by Eq.(5) can be even bigger than 1: the energy is 
coming from optical
fields. Also it is important that this nonlinear transformation is not
dependent on population inversion, so this efficiency can be obtained at room
temperature. 

The efficiency for $\tau\sim 100$ fs can be increased by 
increasing the density of molecules to $10^{17-18}$ cm$^{-3}$. 
For ultra-fast pulsed optical systems ranging in their pulse duration 
from 15 fs to 150 fs, one is able to obtain the same duration for THz
radiation as well. At the same time, using optical pumping beams with 
up to nanosecond pulse duration, one can generate efficiently 
an intense THz pulse of a nanosecond duration. Efficiency of 
generation of THz pulses 
with longer durations is limited by the relaxation of molecular vibrational 
coherence. 
%\section{Current way to produce short THz pulses}

We perform simulations for 
laser pulse duration $\tau = 150$ fs, optical beams 1 and 2 have 
30 $\mu$J with corresponding Rabi frequencies $0.3\cdot 10^{15}$ s$^{-1}$,  
$\wp_v=0.1D$, $\wp_j = 1D$ ($CH_3F$), IR correspond CO$_2$ laser line
9P20, wavelength 500 $\mu$m [$(j=12,K)-(j=11,K)$, K=1-5], 
density $10^{17}$ cm$^{-3}$, 
relaxation rate due to presure broadening is chosen to be $250$ MHz,
detuning $300$ MHz. The
results are shown in Fig.2.

It is worth comparing the estimate above with the currently 
achieved parameters by a variety of methods already considered as
successful for generation of short THz pulses. 
THz pulses with ns durations were achieved in THz semiconductor 
lasers (quantum cascade lasers, p-Ge and n-Si lasers~\cite{qcl,6,7,8,9},
limited by the need for cryogenic cooling). 
In p-Ge lasers \cite{6,7,8}, in the mode-lock regime, 
20 ps-short pulse durations have been achieved 
with peak power up to a few Watts in a few-$\mu$s train of pulses. 
%The n-Si optically pumped lasers produce the ns-long pulses of
%THz radiation \cite{9}.  

Two other optical-pumping-based approaches giving subpicosecond THz
pulses are the photocurrent method using the Auston-switch technique
\cite{10,11,12,13,14} and optical rectification
\cite{13,14}. These methods allow one to generate short THz pulses with
subpicosecond durations, but with rather low intensities.  
%One of the new variants of optical-pumping-based techniques- 
The nonlinear
resonant mixing in semiconductor quantum-well systems, can provide 
short pulses, but also with very low
conversion efficiency, and it works in the mid-IR rather than in THz region
\cite{15}.  

%At the moment, 
The most impressive results in generation of short THz
pulses were achieved by using free electron beam based sources: free-electron
lasers (FEL) \cite{3,4} and synchrotrons \cite{16}. 
%For FELs,
%typical parameters of achieved short THz pulses of radiation are
%-about 1ps duration with 1-40 ns distance between micropulses, grouped
%into a few-$\mu$s trains. The energies in micropulses are about 1-50
%$\mu$J (the parameters of FELBE laser, Rossendorf, Germany). 
%
Recently, the substantially higher power of coherent broadband THz
radiation pulses produced by synchrotron emission were obtained
from the electron beamline \cite{16,17},
up to 100~$\mu$J, hundreds of femtoseconds-short
half-cycle THz pulses. 
%In the pulses of such intensity, the electric filed E
%can reach already about 1 MV/cm. So large electric fields already can displace
%atoms in polar solids, and can open an unique way for investigations of
%structural phase transitions, soft modes, ferroelectricity in solids. The
%magnetic field ($H=E/C$) in these THz pulses also can reach 
%very high levels, of about 0.3~T. 
%So, one can use this transient magnetic field to create magnetic/spin 
%excitations and follow dynamics on ps time scale.

% III One more way....

%\section{Discussion}

%In the current work, we apply the methods of quantum optics
%to the generation of THz pulses in coherently prepared media. We
%decided to use as active media the well-known molecular gases (methanol,
%ethanol, CH$_2$F$_2$, CH$_3$Cl, etc.) that are already used 
%in optically-pumped THz lasers \cite{22,23,24}. 
Our results, obtained in the example of methanol and CH$_3$F 
show that using intensive enough
pumping beams, one can generate the short THz pulses with intensities,
already comparable with intensities of synchrotron or FEL-based THz
sources and with very high efficiency. As was already 
shown above, the proposed
method allows us to go beyond currently reached shortest 
durations of the THz
pulses~\cite{14,16,17,17-1}, i.e. to reach fs-durations. In addition to that, 
in our method,  the duration of 
THz pulse can be controlled by the durations of both the pumping and 
drive beams. 
Besides, it works in either generating single pulses defined by the durations 
of the pump and drive beams, or in generating a sequence of 
relaxation oscillation 
processes-defined pulses if to pump/drive a gas cell by long pulses. 

%One important remark should be made, namely that the phase-matching 
%condition can be strongly modified in the coherently driven media
%so it allows control the direction of THz field generation~\cite{phase-match}.
%Another remark is that the inhomogeneous IR radiation can create 
%a waveguide in the medium in such way that the THz radiation can propagate 
%without diffraction losses~\cite{EIT-waveguide}. 

Moreover, the approach elegantly developed 
in~\cite{harris,sokolov,nazarkin} can be applied to THz 
radiation as well. 
%Namely, the coherence between molecular rotational levels 
%can be generated by optical fields, and then the coherent 
%THz radiation can be scattered from this coherence to generate comb with 
%frequency difference equal to the rotational frequency. 
%Synchronizing all harmonics 
%in the comb would give us a pulse with a duration of
%a period of the THz radiation.  
Strong dispersion of molecular gas in THz region 
of vibrational-rotational resonances create 
additional possibilities~\cite{nazarkin}
to obtain short THz pulses. 
%The possible efficiency of this method will be discussed elsewhere. 

%\section{Conclusion}

In conclusion, we have suggested an efficient method of generation of short 
THz pulses using coherent scattering of drive radiation in 
a coherently prepared atomic or molecular gas. 
%Historically, the molecular gas lasers (f.e., the methanol 
%laser), were the first  laser THz  sources created at 1970s. 
Our work opens up a new way of using molecular gas sources for highly
efficient generation of short intensive THz pulses. 
Applications of the obtained results have a broad range: from molecular
spectroscopy to imaging, monitoring of environment and diagnostics of liquid
and solid materials. 

%\section{Acknowledgments}

We thank A. Chugreev, J. Kono, A. Nazarkin, and O. Portugal
for helpful discussions. We also 
gratefully acknowledge the support from the Office of Naval Research, 
the Air Force Office of Scientific Research, the Defense Advanced Research Projects Agency, and the Robert A.\ Welch Foundation.

%\section*{References}

%\end{document}

\section*{Figure caption}

Figure 1.

Level schemes  for THz generation.
$\W_1$ and $\W_2$ are optical fields in two-photon resonance to excite atomic
or molecular coherence between levels $b$ and $c$. 
$\W_3$ is the IR field, and $\W_4$ is generated THz radiation 
(Double-$\Lambda$ shown by dashed lines, Ladder-$\Lambda$ by solid lines). 
Atomic medium (a). For example, Rb levels are 
$a=5S_{1/2}$, $b=10P_{1/2.3/2}$, $c=6P_{1/2.3/2}$, $d(d')=8D_{3/2,5/2}
(9D_{3/2,5/2})$.  
Molecular medium (b). Level $b$ is the ground state, vibrational quantum
number $v=0$ and angular momentum $j$; $c$ is the level with $v=1$ and $j'$, 
$d$ is the level, $v=0$ and $j+1$; $d'$ is the level $v=1$ and $j'+1$. 
Levels of CH$_3$F shown in (c).

Figure 2.

Results of simulations for double-$\Lambda$ scheme in CH$_3$F for two temporal
scenarios. 
Scenario 1. 
Fields $\W_1$ and $\W_2$ and prepared by them coherence $\rho_{bc}$ 
are shown in a).  Probe IR field $\W_3$ and generated THz field $\W_4$ are 
shown in b).  Efficiency of nonlinear transformation is shown in c).  
Scenario 2.
Fields $\W_1$ and $\W_2$ in STIRAP configuration 
and prepared by them coherence $\rho_{bc}$ 
are shown in d).  Probe IR field $\W_3$ and generated THz field $\W_4$ are 
shown in e).
Parameters used in simulations are in the text.

\newpage

\vspace*{5cm} Figure 1, N.G. Kalugin, et al., 
``Generation of strong short coherent terahertz pulses in 
a ladder-Lambda system'' 

\newpage

\vspace*{5cm} Figure 2, N.G. Kalugin, et al., 
``Generation of strong short coherent terahertz pulses in 
a ladder-Lambda system''

\end{document}